\documentclass[%
 aip,
 amsmath,amssymb,
 reprint,%
]{revtex4-1}

\usepackage{graphicx}% Include figure files
\usepackage{dcolumn}% Align table columns on decimal point
\usepackage{bm}% bold math
\usepackage[utf8]{inputenc}
\usepackage[T1]{fontenc}
\usepackage{mathptmx}
\usepackage{etoolbox}
\usepackage{amsmath}  %To have text in math $$ to not be in italics "use \text{normal text}"
\usepackage{upgreek}  %non-italic greek letters  "put up before the name" ex. \upmu
\usepackage{gensymb}
\usepackage[super]{nth}  %to add 0th / 1st etc

\makeatletter
\def\@email#1#2{%
 \endgroup
 \patchcmd{\titleblock@produce}
  {\frontmatter@RRAPformat}
  {\frontmatter@RRAPformat{\produce@RRAP{*#1\href{mailto:#2}{#2}}}\frontmatter@RRAPformat}
  {}{}
}%
\makeatother
\begin{document}

\preprint{AIP/123-QED}

\title{Density dependence of the excitation gaps in an undoped Si/SiGe double-quantum-well heterostructure}
% Force line breaks with \\
\author{D. Chen}
 %\email{dclover23@ufl.edu.}
\affiliation{ 
Department of Physics, University of Florida, Gainesville, FL 32611, USA%\\This line break forced with \textbackslash\textbackslash
}%
\author{S. Cai}%

\affiliation{ 
Department of Physics, University of Florida, Gainesville, FL 32611, USA%\\This line break forced with \textbackslash\textbackslash
}%

\author{N.-W. Hsu}

\affiliation{%
Department of Electrical Engineering and Graduate Institute of Electronic Engineering, National Taiwan University, Taipei 10617, Taiwan%\\This line break forced% with \\
}%

\author{S.-H. Huang}

\affiliation{%
Department of Electrical Engineering and Graduate Institute of Electronic Engineering, National Taiwan University, Taipei 10617, Taiwan%\\This line break forced% with \\
}%

\author{Y. Chuang}

\affiliation{%
Department of Electrical Engineering and Graduate Institute of Electronic Engineering, National Taiwan University, Taipei 10617, Taiwan%\\This line break forced% with \\
}%

\author{E. Nielsen}

\affiliation{%
Sandia National Laboratories, Albuquerque, New Mexico 87185, USA%\\This line break forced% with \\
}%

\author{J.-Y. Li}
 %\homepage{http://www.Second.institution.edu/~Charlie.Author.}
\affiliation{%
Department of Electrical Engineering and Graduate Institute of Electronic Engineering, National Taiwan University, Taipei 10617, Taiwan%\\This line break forced% with \\
}%
\affiliation{%
Taiwan Semiconductor Research Institute, Hsinchu 30078, Taiwan%\\This line break forced% with \\
}%

\author{C. W. Liu}

\affiliation{%
Department of Electrical Engineering and Graduate Institute of Electronic Engineering, National Taiwan University, Taipei 10617, Taiwan%\\This line break forced% with \\
}%

\author{T. M. Lu}

\affiliation{%
Sandia National Laboratories, Albuquerque, New Mexico 87185, USA%\\This line break forced% with \\
}%

\author{D. Laroche}%
    \altaffiliation{\textbf{Email of Author to whom correspondence should be addressed:} dlaroc10@ufl.edu}
\affiliation{ 
Department of Physics, University of Florida, Gainesville, FL 32611, USA%\\This line break forced with \textbackslash\textbackslash
}%

\date{\today}% It is always \today, today,

\begin{abstract}
We report low-temperature magneto-transport measurements of an undoped Si/SiGe asymmetric double quantum well heterostructure. The density in both layers is tuned independently utilizing a top and a bottom gate, allowing the investigation of quantum wells at both imbalanced and matched densities. Integer quantum Hall states at total filling factor $\nu_{\text{T}} = 1$ and $\nu_{\text{T}} = 2$ are observed in both density regimes, and the evolution of their excitation gaps is reported as a function of density. The $\nu_{\text{T}} = 1$ gap evolution departs from the behavior generally observed for valley splitting in the single layer regime. Furthermore, by comparing the $\nu_{\text{T}} = 2$ gap to the single particle tunneling energy, $\Delta_{\text{SAS}}$, obtained from Schr\"{o}dinger-Poisson (SP) simulations, evidence for the onset of spontaneous inter-layer coherence (SIC) is observed for a relative filling fraction imbalance smaller than ${\sim}50\%$.
\end{abstract}

\maketitle

%Intro take 2 (first 2 paragraphs):
% Exotic behavior such as novel fractional quantum Hall states /cite{eisenstein_1992, suen_1992, hamilton_1996} and exciton condensation \cite{murphy_1994, kellogg_2002, tutuc_2004, tiemann_2008, li_2017} systems have been shown to arise in bilayer systems where intralayer interactions are comparable or stronger to interlayer interactions.  These low-disorder quantum phenomenon have been predominently observed in GaAs heterostrucres and graphene.  

%  In recent years, the substantial low-temperature mobility improvement achieved in SiGe-based heterostructures \cite{Lu papers} has enabled the study of single-layer low-disorder phenomenon, such as the fractional quantum hall \cite{...} effect, valleytronics \cite{valleytronics} and long lifetime and high fidelity electron spin qubits and quantum dots\cite{Scappucci, Eriksson, Peta, maybe 1-2 other} in silicon-based heterostructures.  Extending such studies to coupled Si/SiGe bilayers \cite{Laroche_2015} would provide a new and readily scalable platform for the study of indirect exciton condensation and its potential applications\cite{resitaneless transistors, quantum hall droplets}.  Studying the evolution of valley splitting in Si/SiGe bilayers is also of crucial interest for its potential to realize long-lived silicon spin qubits \cite{goswami_2007}. 

Owing to their compatibility with the widespread complementary metal-oxide-semiconductor (CMOS) technology enabling a scalable and affordable fabrication process, Si-based heterostructures continue to attract a lot of attention for both fundamental and applied research.\cite{ng_2002} In particular, the substantial low-temperature mobility improvements achieved in SiGe-based heterostructures \cite{lu_2009} have enabled various studies of low-disorder phenomena, such as the fractional quantum hall effect,\cite{lu_2012} valley splitting \cite{weitz_1996,neyens_2018,mcjunkin_2021}, quantum dots\cite{xue_2019, mills_2019,lawrie_2020},  as well as long lifetime and high fidelity electron spin qubits\cite{simmons_2011, russ_2018} that are predominantly studied alongside GaAs-\cite{petta_2005, sarma_2008} and graphene-\cite{bolotin_2009, shimazaki_2015}based devices. 

The vast majority of the experimental efforts on quantum transport in SiGe-based heterostructures  have focused on the development and characterization of single layer devices, while coupled Si/SiGe bilayers \cite{malissa_2006,laroche_2015} have received little attention.  However, bilayer systems present a promising platform in their own right for studying  interlayer coherence and the onset of indirect exciton condensation.\cite{eisenstein_1992,tutuc_2004,tiemann_2008,burg_2017,li_2017} In addition, the evolution of valley splitting in Si/SiGe bilayers is of crucial interest for the realization of long-lived silicon spin qubits.\cite{goswami_2007} Undoped Si/SiGe double quantum wells have only been investigated in top-gated devices,\cite{laroche_2015} severely limiting the density tunability of the devices' bottom well. Unlike these prior studies, the device presented in this letter hosts a bilayer system that is dual-gated, allowing independent control of the density in both layers simultaneously. Utilizing this enhanced control, the evolution of the density dependence of the excitation gaps at total filling fractions $\nu_{\text{T}}=1$ and $\nu_{\text{T}}=2$ is reported both at mismatched and matched densities.

\begin{figure*}[t]
\includegraphics[width=1.75\columnwidth]{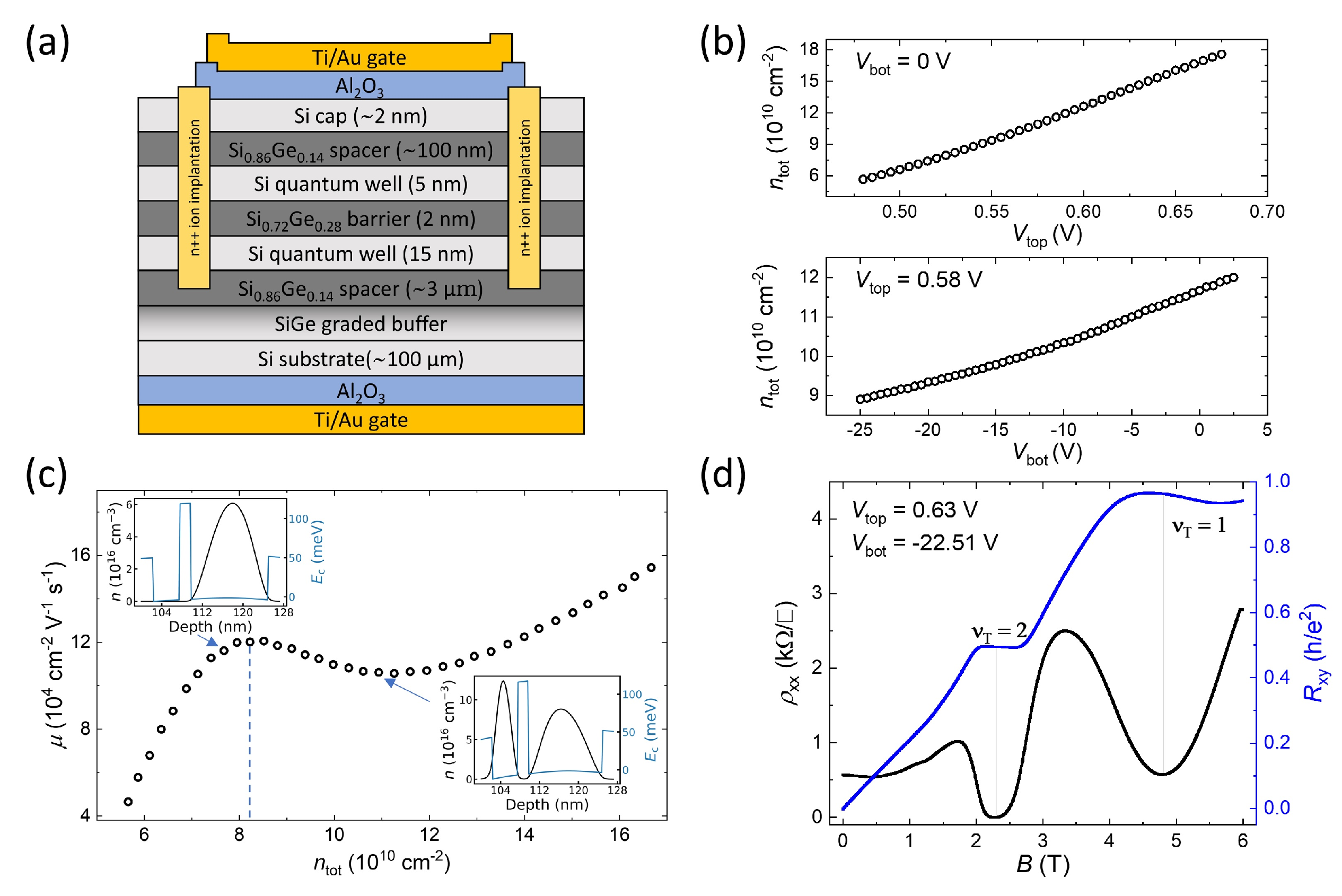} %fig_1v6.jpg
\caption{\label{fig:device_char}(a) Complete layer schematic of the device. (b) Top Panel: Total density dependence vs top gate voltage with the bottom gate voltage set to 0. Bottom Panel: Total density dependence vs bottom gate voltage with a top gate voltage of 0.58 V. (c) Total 2D density vs mobility dependence at fixed $V_{\text{bot}} = 0$ V. A similar plot over a wider density range is shown in the supplementary material. Top Left (Right) inset: SP simulation results describing the band diagram and 3D electron density distribution below (above) $n_{\text{crossover}}$, the blue dotted line in the main panel.  (d) Hall resistance (blue) and longitudinal resistivity (black) as a function of magnetic field for $V_{\text{top}}$ = 0.63 V and $V_{\text{bot}}$ = $-22.51$ V. The top and bottom wells' densities are $5.41\times10^{10}\ \text{cm}^{-2}$ and $5.82\times10^{10}\ \text{cm}^{-2}$, respectively.  The method to extract each layers' densities is detailed in the supplementary material.}
\end{figure*}

Closely following previous heterostructure designs,\cite{laroche_2015} the double-quantum well structure presented in this letter was designed to contain a $5\ \text{nm}$ top quantum well, a $2\ \text{nm}$ $\text{Si}_{\text{0.72}}\text{Ge}_{\text{0.28}}$ interwell barrier and a $15\ \text{nm}$ bottom quantum well. In contrast to the previous structure, the interwell barrier's Ge content was increased to 28\% to suppress tunnelling between the two wells. The heterostructure was grown in an ultra-high-vacuum chemical vapor-deposition system with Si$\text{H}_{\text{4}}$ and Ge$\text{H}_{\text{4}}$ as precursors on a p-type Si substrate. As depicted in Fig.~\ref{fig:device_char}(a), the layer composition is, from bottom to top: a graded SiGe virtual substrate with a maximal Ge composition of 14$\%$, a ${\sim}3\ \upmu\textrm{m}$ thick relaxed $\text{Si}/\text{Si}_{\text{0.86}}\text{Ge}_{\text{0.14}}$ spacer layer, the (strained) double quantum well structure described above, an additional ${\sim}100\ \upmu\textrm{m}$ $\text{Si}/\text{Si}_{\text{0.86}}\text{Ge}_{\text{0.14}}$ spacer layer and a $2\ \text{nm}$ Si cap.

The asymmetric quantum well structure ensures that, in the uncoupled limit, the ground state energies in the two wells are different.  Thus, during top gate operations, a sizable electron density can accumulate in the wider bottom layer before electrons start to populate the top layer, drastically reducing further electron accumulation in the bottom layer through screening. Such a design is necessary owing to the large separation, ${\sim}100\ \upmu\textrm{m}$, between the bottom gate and the bottom quantum well, limiting the degree of density tuning achievable from the bottom gate alone. 

To enable electrical measurements on the bilayer, a heterostructure field-effect transistor (HFET) was patterned following standard fabrication processes described in detail in a previous letter.\cite{laroche_2015} First, heavy phosphorus ion implantation was used for ohmic contacts. Prior to the deposition of the Hall bar-shaped Ti/Au gate, 1000 cycles of atomic-layer-deposited $\text{Al}_{2}\text{O}_{3}$ were grown. Finally, the back Ti/Au gate was implemented by mechanically lapping the substrate down to ${\sim}100\ \upmu \text{m}$, followed by a similar oxide and metal deposition scheme.

The device was tested in a cryo-free Blue Fors dilution refrigerator with a base temperature of $T\,{\sim}\,10\ \text{mK}$. Standard low-frequency (37.3 Hz) lock-in measurement techniques were used in a four-terminal geometry with a constant 50 nA excitation current. Prior to the initial operation of the device, a large negative bias ($-7$ V) was applied to evacuate negative charges trapped at the dielectric layer interface, improving device reliability. The measurements were found to be consistent over several thermal cycles.

The Hall density of the device is extracted from the slope of the Hall resistance $(\text{R}_{\text{xy}})$ at small magnetic fields ($B \leq 0.1\,\text{T}$), and its dependence on top (bottom) gate voltage at $V_{\text{bot}} = 0$ V ($V_{\text{top}} = 0.58$ V) is shown in the top (bottom) panel of Fig.~\ref{fig:device_char}(b).  Both curves show near-linear behavior, while the bottom gate voltage dependence has a smaller slope owing to a larger gate-to-well separation.

Total Hall densities were tuned from $6.8\times10^{10}\ \text{cm}^{-2}$ to $34.2\times10^{10}\ \text{cm}^{-2}$, with a maximal Hall mobility of $30.8 \times 10^{4}\ \text{cm}^{2}/ \text{(V} \cdot \text{s)}$ achieved at the highest density.  The crossover density (the electronic density required to start populating the top quantum well) was observed at $n_{\text{crossover}} = 8.22\times10^{10}\ \text{cm}^{-2}$.  Identification of $n_{\text{crossover}}$ and confirmation of bilayer behavior is assessed from the mobility drop in the density versus mobility plot, caused by inter-layer scattering,\cite{stormer_1982} as exemplified in the main panel of Fig.~\ref{fig:device_char}(c).

%As mentioned previously, an undoped heterostructure is used in order to circumvent the difficulties associated with using a modulation-doping scheme to grow a bilayer Si/SiGe structure. Thus, a gate is used to capacitively induce charge carriers in the bilayer. The independent population of both quantum wells simultaneously is achieved by using the novel dual-gating scheme from both the top and bottom side of the heterostructure. 

%remove Electrons do not accumulate at the oxide/semiconductor interface at low voltage bias and low temperature due to the charge distribution never reaching thermal equilibrium \cite{T.M.Lu_2011}. In order to account for this non-equilibrium behavior within the SP simulation, the bandgap of the upper and lower half of the SiGe's top and bottom spacer was artificially increased in order to prevent electrons from accumulating at the oxide/semiconductor interfaces. (no need to include this b/c it detracts from the message)   Maybe strip this to only one sentence and cite.

The experimental crossover density was also replicated using an iterative, self-consistent SP simulation. Using the nominal growth parameters and thicknesses, the experimental crossover density was reproduced via the simulation by shifting the lower boundary condition, corresponding to a constant offset in the bottom gate voltage. The insets of Fig.~\ref{fig:device_char}(c) illustrate the simulated band diagram and electron density distribution before and after $n_{\text{crossover}}$.  To prevent charge accumulation at either interface, the conduction band of the upper (lower) half of the heterostructure's top (bottom) spacer is artificially increased to be 500 meV above that of the quantum wells.\cite{laroche_2015}  We also note that it was possible to reproduce the experimental crossover density without utilizing a bottom gate voltage offset by considering a modified structure with a top well width of $3.8\ \text{nm}$, an interlayer barrier of $2.2\ \text{nm}$ and a bottom well width of $15\ \text{nm}$.

%my figures are the ones below this
%Fig.~\ref{fig:fig_2}%
\begin{figure*}[t]
\includegraphics[width=1.5\columnwidth]{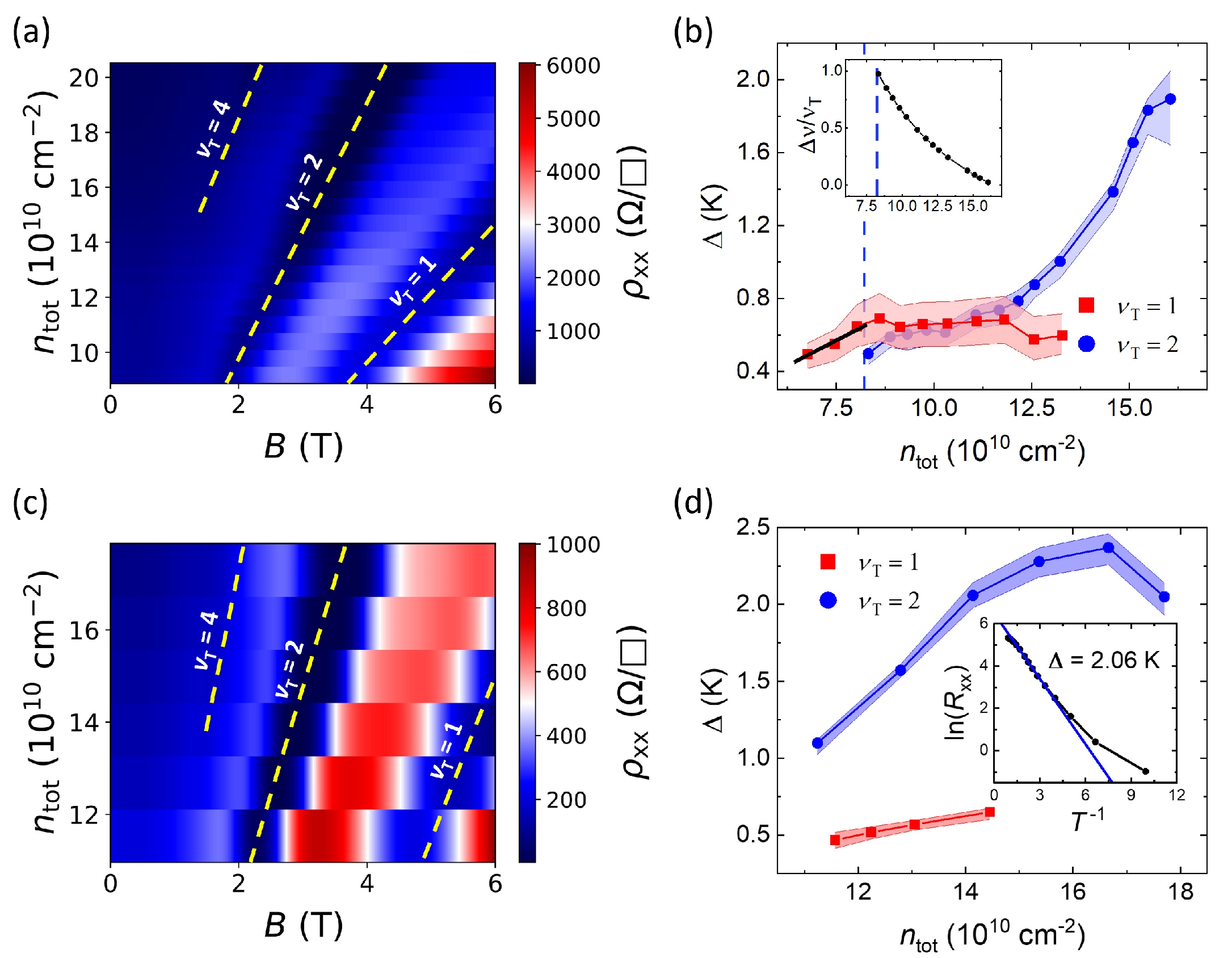} %fig_2_comb_v10.jpg
\caption{\label{fig:exp_data} (a) 2D color map exhibiting the longitudinal resistivity, $\rho_{\text{xx}}$ as a function of $n_{\text{tot}}$ and $B$ for variable imbalanced densities.  The matched density case is shown in panel (c).  (b) Density dependence of the extracted activation gaps at variable imbalanced densities with confidence intervals for $\nu_{\text{T}} = 1$ (red squares) and $\nu_{\text{T}} = 2$ (blue circles). $n_{\text{crossover}}$ is indicated by the blue dotted line. In addition, a black line indicates the points used to fit the estimated disorder broadening, $\Gamma$, and the linear valley splitting coefficient, $c_{\text{B}}$. The matched density case is shown in panel (d).  Inset (b): Relative filling fraction imbalance as a function of total density (same scale as main graph) for the variable imbalanced density case. As the bottom well’s density marginally increases beyond the crossover density, the top layer’s density slowly catches up to the bottom layer’s density, thereby reducing the filling fraction imbalance. Inset (d): Temperature dependence of the $\rho_{\text{xx}}$ minima in an Arrhenius plot for $\nu_{\text{T}} = 2$ at $V_\text{top} = 0.64$ V and $V_\text{bot} = -12.368$ V in the matched density regime.}
\end{figure*}

Two density regimes are studied in this letter. The variable imbalanced density regime describes experimental conditions where only $V_{\text{top}}$ is increased from 0 to positive values, while $V_{\text{bot}}$ is fixed at 0 V, as exemplified in Fig.~\ref{fig:device_char}(c). Beyond $n_{\text{crossover}}$, the density in the top layer steadily increases towards matched densities, while the density in the bottom layer remains nearly constant, changing by less than 4$\%$ according to simulations. In contrast, both gates are biased in the matched density regime, yielding densities within 6$\%$ of one another in each layer, under the assumption of ideal gate screening by the 2DEGs.  Independent contacts to both layers or a capacitive measurement scheme \cite{dorozhkin_2018} would be needed to further refine this estimate. Details on the tuning scheme can be found in the supplementary material. An instance of this later case is shown in Fig.~\ref{fig:device_char}(d); where the traces of the Hall resistance, $R_{\text{xy}}$, and of the longitudinal resistivity, $\rho_{\text{xx}}$, at a front gate voltage of $V_{\text{top}} = 0.63$ V and a back gate voltage of $V_{\text{bot}} = -22.51$ V are presented. In both regimes, the data clearly exhibits quantum Hall plateaus developing at $\nu_{\text{T}}=1$ and $\nu_{\text{T}}=2$, within $3\%$ of $h/e^2$.

The evolution of $\rho_{\text{xx}}$ as a function of magnetic field and total density is presented in Fig.~\ref{fig:exp_data} for both variable imbalanced (a) and matched (c) densities.  Quantum Hall plateaus are clearly observed for $\nu_{\text{T}}=1$ and $\nu_{\text{T}}=2$, indicated by dotted yellow lines. Although less developed, the $\nu_{\text{T}}=4$ minima is also observed. 

The main results of this letter are presented in Fig.~\ref{fig:exp_data}(b) and Fig.~\ref{fig:exp_data}(d) where the quantum Hall excitation gaps, extracted from the temperature dependence of the $\rho_{xx}$ minima, are shown for the variable imbalanced and matched density regimes, respectively.  The inset in Fig.~\ref{fig:exp_data}(d) shows a typical example of the dependence $\uprho_{xx}$ has on the temperature in an Arrhenius plot. The linear part of the plot was systematically determined and fitted to an activation model,\cite{boebinger_1990} and the extracted slope was used to determine the excitation gap $\Delta$: $\rho_{\text{xx}}(T) = \rho_{0} \text{e}^{-\Delta/2T}$. The deviation from linearity at low temperatures is attributed to interstitial hopping.\cite{ebert_1983} The shaded confidence intervals in the main panel are determined from either adding and/or subtracting an additional point to/from the determined linear activation interval, whenever possible.  For additional details on the fitting procedure, see the supplementary material.

%MAKE SURE TO CHANGE ESTIMATES
%add a sentence for fully independent layers vs bilayers
In the single layer regime, strained Si contains two degeneracies per Landau Level: spins $(\uparrow\downarrow)$ and valleys (+/-). For $n_{\text{tot}} = 10 \times10^{10}\ \text{cm}^{-2}$, the single layer valley splitting is estimated to be ${\sim}1.35\ \text{K}$ at 4.13 T.\cite{wuetz_2020} At the same magnetic field, the spin gap would be ${\sim}2.79\ \text{K}$ without considering many-body effects.  This implies that the $\nu_{\text{T}} = 1$ ($\nu_{\text{T}} = 2$) state is associated with the valley (spin) degree of freedom.  Upon entering the bilayer regime, an additional degeneracy, the layer degree of freedom (S/AS), comes into play.  The tunneling gap, $\Delta_{\text{SAS}}$, \cite{boebinger_1990} characterizing this degeneracy can be estimated from the SP simulation. As we do not observe Landau level crossing in our magneto-transport measurements (Fig.~\ref{fig:exp_data}(a)), we can safely attribute the bilayer $\nu_{\text{T}} = 1$ state to valley splitting. The $\nu_{\text{T}} = 2$ state is attributed either to spin-splitting or to interlayer effects. In the latter case, two mechanisms can give rise to the quantum Hall state: a single-particle state caused by interlayer tunnelling \cite{boebinger_1990, he_1993} or a many-body state resulting from spontaneous interlayer coherence (SIC). \cite{ he_1993,murphy_1994,sawada_1998}

%Talking about g factor (valley splitting) for nu = 1
Although valley splitting in vertically coupled Si/SiGe bilayers has not been investigated yet, the single layer regime has attracted a lot of interest; both  experimentally\cite{lai_2004,goswami_2007} and theoretically.\cite{friesen_2006,friesen_2007} For single layers, the conduction band valley degeneracy is lifted by a sharp quantum well interface.  Atomic-scale disorder, attributed to random SiGe alloys and steps in the quantum well, causes destructive interference that suppresses valley splitting. Theoretical calculations predict a linear dependence that extrapolates to a value that coincides with sample-dependent Landau level disorder broadening at zero density.\cite{friesen_2007}

The evolution of the $\nu_{\text{T}}=1$ gap in the single layer regime, presented in red in Fig.~\ref{fig:exp_data}(b), follows these expectations. Linearly fitting the data below $n_{\text{crossover}}$, we obtain an estimate for the disorder broadening $\Gamma\,{\sim}\,0.327$ K and for a linear coefficient $c_{\text{B}}\,{\sim}\,0.29$ $\text{K}/\text{T}$, consistent with previous studies in Si/SiGe heterostructures reporting similar results:\cite{wuetz_2020} $\Gamma\,{\sim}\,0.435$ K and $c_{\text{B}}\,{\sim}\,0.326$ $\text{K}/\text{T}$. However, this behavior changes drastically in the bilayer regime. The single layer linear dependence strongly decreases and nearly flattens out.  This remains true in the matched density case where the linear coefficient is nearly reduced by a factor of 2 to $c_{\text{B}}\,{\sim}\,0.15$ $\text{K}/\text{T}$. Further theoretical and experimental studies of this unexpected behavior are required to determine whether Si bilayers exhibit large valley splitting at low magnetic fields or even spontaneous valley polarization.\cite{friesen_2006,hossain_2020}

%Fig.~\ref{fig:fig_3}%
\begin{figure}
\includegraphics[width=1\columnwidth]{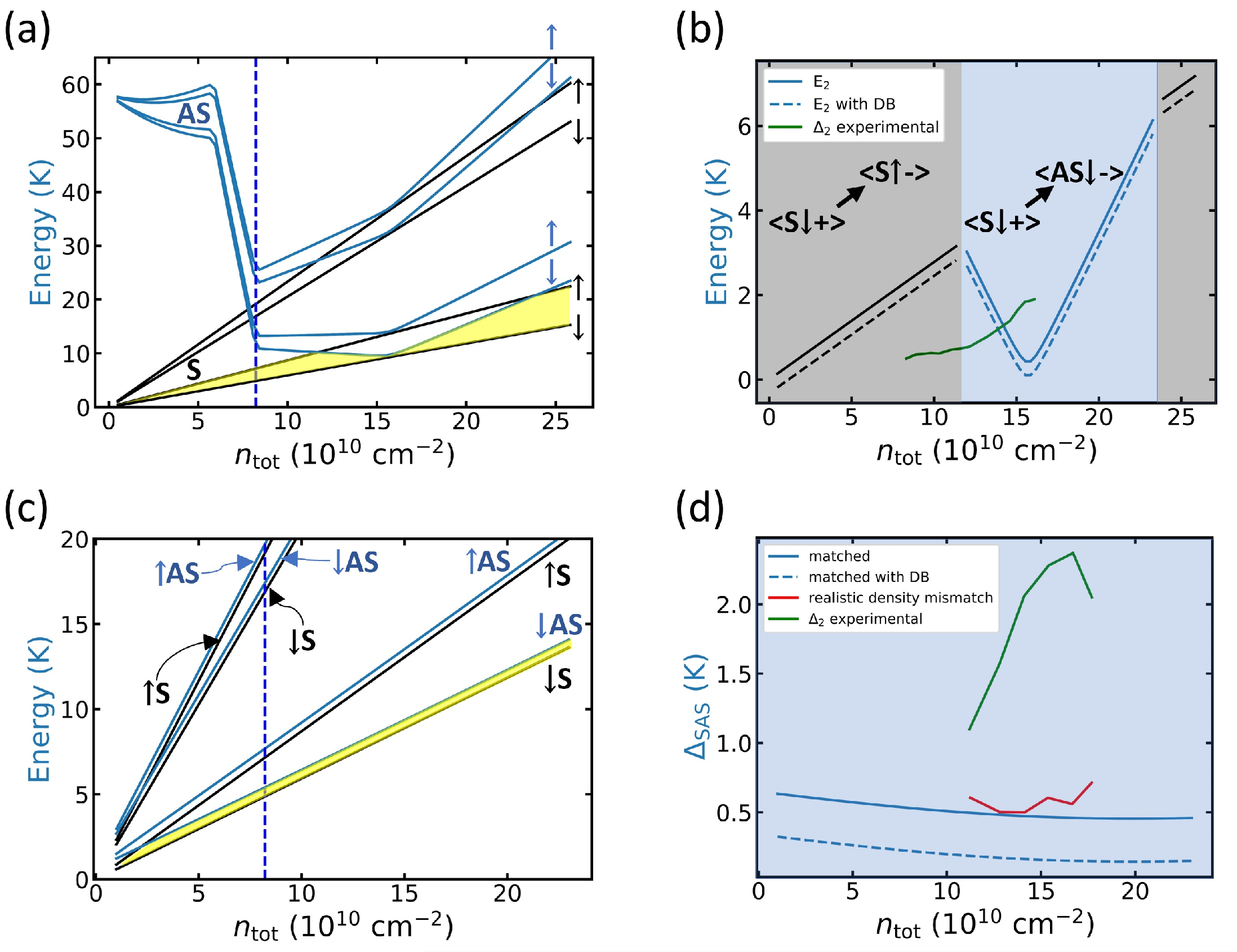} %fig3_v12.jpg
\caption{\label{fig:sp_sims} (a) Simulated energy fan diagram depicting the \nth{0} and \nth{1} Landau levels for variable imbalanced densities. The matched density case is shown in panel (c). The symmetric (S) state subband is shown as black lines while the asymmetric (AS) state subband is shown as blue lines. Spin is denoted by arrows. The dotted line indicates where $n_{\text{crossover}}$ occurs. Valley splitting is omitted from the calculations. The yellow shaded regions indicate the energy gaps considered for panel (b) and (d). (b) As a function of density, $\text{E}_{\text{2}}$ can either result from a spin state transition (shaded black) or from a layer state transition (shaded blue). (d) $\Delta_{\text{SAS}}$ values for exactly matched densities (blue) and for the slightly mismatched densities measured experimentally (red); see Table S.I. in the supplementary material. The dotted lines in panels (b) and (d) take into consideration disorder broadening. The experimentally measured $\nu_{\text{T}} = 2$ excitation gaps are shown as green lines in panels (b) and (d).}
\end{figure}

%talking about interlayer affects for nu = 2
To understand the evolution of the $\nu_{\text{T}} = 2$ state, one has to consider the interplay between spin-splitting and interlayer effects.  One interlayer mechanism that may induce the $\nu_{\text{T}} = 2$ quantum Hall state in coherent bilayers\cite{eisenstein_1992} occurs when the tunneling between the two quantum wells is sufficiently strong, resulting in an overlap of the wave functions in each well. As a result, the individual layer eigenstates hybridize into symmetric and antisymmetric states with single-particle tunnel splitting gap, $\Delta_{\text{SAS}} = \text{E}_{\text{AS}} - \text{E}_{\text{S}}$.\cite{hu_1992}  $\Delta_{\text{SAS}}$ depends strongly on the height and width of the interwell barrier.
%Using SP simulations, this dependence has been calculated and plotted for the mismatched (matched) density case in Fig 4 (b)/(d).    %go into SP now and compare

%talk about matched non monotonic/doesnt extrapolate to 0/comment about evolution of delta sas w/ the field
Spontaneous interlayer coherence (SIC) is another interlayer mechanism that can give rise to the $\nu_{\text{T}} = 2$ state. As a signature of exciton condensation between spatially separated electrons and holes, SIC induces a coherent state that can occur solely due to Coulomb interactions in the limit of low interlayer tunneling; when the ratio between interlayer and intralayer Coulomb interactions is less than a critical value: $d/l_{\text{b}} \lesssim 1.8$.\cite{spielman_2000,kellogg_2002} Here, $d$ is the distance between the centers of each quantum well and $l_{\text{b}} = (\hbar/eB)^{1/2}$ is the magnetic length.  The device reported in this letter has $d/l_{\text{b}}$ values ranging between $0.62$ and $1.24$, well within the range where SIC can be observed. Furthermore, the ratio of the SP simulated $\Delta_{\text{SAS}}$ gap to the Coulomb energy ($e^2/\epsilon l_{\text{b}}$) is ${\sim} 5 \times 10^{-4}$, indicating that the device is indeed in the Coulomb dominated regime. 

%gen nu = 2 is for SIC when measured activation gap is > than deltasas and does not follow the same dependence. (find papers for this)

%Both spin (black line) and $\Delta_{\text{SAS}}$ (blue line) sub-levels are presented.

Both spin-splitting and interlayer tunnelling are captured within the SP simulations. In addition to the two different cases used to reproduce $n_{\text{crossover}}$, two different interwell barrier heights, $\phi_{\text{b}}$, were used. The first one corresponds to a Ge content of 28.8$\textrm{\%}$, giving $\phi_{\text{b}} = 141.2$ meV; while the other assumes a 25.8$\%$ Ge content, giving $\phi_{\text{b}} = 122.8$ meV.\cite{schaffler_1997}  Together, we have thus simulated four cases. We note that further increasing the Ge content of the barrier would have negligible effect on its height.  Only the case with as-grown parameters and higher interwell barrier is shown in Fig.~\ref{fig:sp_sims}. The other three cases are shown in the supplementary material. The general shape of the density dependence of $\Delta_{\text{SAS}}$ is nearly identical in all four cases. However, the magnitude of $\Delta_{\text{SAS}}$ can be up to ${\sim}96\%$ larger for the case of thinner top well width and weaker interlayer barrier, as opposed to as-grown parameters and higher interwell barrier. Panel (a) (Panel (c)) of Fig.~\ref{fig:sp_sims} depicts the evolution of the energy fan diagrams for the variable imbalanced (matched) density regime. Here, only spin splitting and $\Delta_{\text{SAS}}$ are considered while valley splitting is omitted from the calculations, resulting in 4 energy sub-levels per Landau level. As such, the simulation strictly provides information about $\text{E}_{\text{2n}}$, the energy of the even quantum Hall states. The yellow shaded regions denote the simulated $\text{E}_{2}$. The simulated second Landau energy level, $\text{E}_{2}$, for the variable imbalanced density case is also shown in Fig.~\ref{fig:sp_sims}(b) while the simulated $\Delta_{\text{SAS}}$ for the matched density case is shown in Fig.~\ref{fig:sp_sims}(d).  

%transition between two linear regimes: one in single layer one in bilayer (nu = 2 non-monotonic) stark diff for matched, both are non-mon & magnitudes switch
%introduce layer imbalance estimate (champagne) 

%now compare extracted gaps to SP simulations

From theory,\cite{wuetz_2020} we expect the experimental gap to be $\Delta_{2} = \text{E}_{2} - \text{E}_{1} - \Gamma = \text{E}_{2} - \Delta_{1}  - 2\Gamma$. In the variable imbalanced density case at low density, the simulated $\text{E}_{2}$ and the experimentally measured $\Delta_{2}$ are similar. Starting in the single layer regime, a linear dependence is observed as long as $\text{E}_{2}$ originates from a spin state transition.  A downturn in the gap magnitude is then observed with increasing density once the state originates from $\Delta_{\text{SAS}}$.  However, the measured magnitude of $\Delta_{2}$ is notably smaller than the simulated $\text{E}_{2}$.  In addition, the theoretical downturn in $\text{E}_{2}$ is reduced to the point of flattening, and its experimental onset occurs at a reduced density, $9.8\times10^{10}\ \text{cm}^{-2}$.  As shown in Fig.~\ref{fig:sp_sims}(b), disorder broadening alone cannot explain the observed discrepancy.  A non-linearity of the valley splitting excitation gap, $\Delta_{1}$, might explain this discrepancy, but the same non-linearity prevents us from extrapolating the experimentally measured $\Delta_{1}$ to lower densities and magnetic fields. 

At higher densities, the behavior of $\Delta_{2}$ departs more strongly from the simulated $\text{E}_{2}$, shown in Fig.~\ref{fig:sp_sims}(b).  The downturn is rapidly inverted and the overall magnitude of $\Delta_{2}$ overtakes that of the disorder-broadening corrected $\Delta_{\text{SAS}}$ as the layers approach matched densities, where $\Delta_{\text{SAS}}$ is calculated to be at its weakest.  Both the qualitative departure from the $\Delta_{\text{SAS}}$ density dependence as well as the increased $\Delta_{2}$ value are consistent with the onset of SIC.\cite{murphy_1994}  This is further confirmed by the behavior of $\Delta_{2}$ in the density matched case (Fig.~\ref{fig:exp_data}(d)). Here, $\Delta_{2}$ has a concavity opposite to that of $\Delta_{\text{SAS}}$ (Fig.~\ref{fig:sp_sims}(d)) and a magnitude of up to 10 times larger than that of the disorder-broadened $\Delta_{\text{SAS}}$, at $\Delta_{2} = 2.37$ K. Overall, the SP simulations show that neither $\Delta_{\text{SAS}}$, spin-splitting nor Landau level spacing alone reproduces the qualitative behavior obtained experimentally.  Extrapolating the density dependence of $\Delta_{1}$ to lower magnetic fields is also insufficient to account for the differences between the measured $\Delta_{2}$ and the simulated $\Delta_{\text{SAS}}$.  

Considering the density imbalance between the layers, shown in the inset of Fig.~\ref{fig:exp_data}(b), the onset of $\Delta_{2}$'s magnitude increase coincides with a relative density imbalance $\frac{\Delta \nu}{\nu_{\text{T}}} = \frac{\nu_1 - \nu_2}{2} \approx 0.5$ at $n_{\text{tot}} = 11.07\times10^{10}\ \text{cm}^{-2}$. Prior studies in GaAs bilayers\cite{spielman_2004, champagne_2008} have reported that SIC remains intact for relative density imbalances as large as $\frac{\Delta \nu}{\nu_{\text{T}}} = 0.5$, which is in quantitative agreement with our observations.   

Three main parameters influence the strength of $\Delta_{2}$ induced by SIC: the ratio between interlayer and intralayer Coulomb interactions $d/l_{\text{b}}$, the tunneling strength $\Delta_{\text{SAS}}$ and the density imbalance $\Delta \nu$.  As the total density is increased in our device, $d/l_{\text{b}}$ increases while $\Delta_{\text{SAS}}$ decreases. $\Delta \nu$ also decreases for the density imbalanced regime.  While the excitation gap has previously been shown to increase with decreasing $d/l_{\text{b}}$\cite{he_1993, murphy_1994} and increasing $\Delta \nu$,\cite{tutuc_2003, clarke_2005, wiersma_2006, champagne_2008} tunnelling has been shown to weaken the SIC induced quantum state.\cite{he_1993, lay_1994}  Therefore, we attribute the increase in $\Delta_{2}$ with increasing density to the reduced tunnelling.  Indeed, going towards higher density decreases tunneling as the wave functions are pushed farther apart. The downturn of $\Delta_{2}$ at the highest densities in the matched density case coincides with the saturation of the simulated $\Delta_{\text{SAS}}$ (Fig.~\ref{fig:sp_sims}(d)) and can be explained by the increased $d/l_{\text{b}}$ at constant tunnelling. These effects are likely noticeable in our device since tunnelling ($\Delta_{\text{SAS}} \geq 500$ mK) is over 1000 times larger than prior studies probing similar effects.\cite{tutuc_2003, clarke_2005, zhang_2013} While our observations are consistent with a SIC-induced $\Delta_{2}$, we cannot fully rule out that a significant reduction of valley splitting at lower magnetic fields is at the origin of this phenomenon since the value of $\Delta_{1}$ at $B \sim 2$ T is neither know experimentally nor theoretically.        

%It is also worth noting that the density mismatched curve converges towards $\Delta_{2}\,{\sim}\,2$ K as the top gate is tuned to induce identical density in both layers, nearly coinciding with the matched density curve, for $\nu_{\text{T}} = 2$.  

% The downturn of the $\Delta_2$ gap does occur later in the SP simulation compared to experiment. This can be attributed to possible width dependence of the electron Land\'{e} g factor that has been shown to increase as the width is lowered for GaAs. \cite{knap_1999} With a higher Land\'{e} g factor of about 5, the SP simulation bring the downturn more inline with experimental values.

%Thus, SIC likely occurs for matched layers; while $\Delta_{\text{SAS}}$ is likely attributed to mismatched layers due to the large mismatched and the enhanced intrinsic $\Delta_{\text{SAS}}$ gap.

In conclusion, we have fabricated and performed magneto-transport measurements on a novel back-gated Si/SiGe bilayer system. The density dependence of the excitation gaps for $\nu_{\text{T}} = 1$ and $\nu_{\text{T}} = 2$ has been studied in both the variable imbalanced and matched density regimes and has been compared to values obtained from SP simulations. The $\nu_{\text{T}} = 1$ state is attributed to valley-splitting while the $\nu_{\text{T}} = 2$ state is attributed to spin-splitting at low density and interlayer effects at larger density.  Stark departure in the density evolution of the $\nu_{\text{T}} = 1$ excitation gap between the single and the bilayer regimes has been reported, highlighting the need for theoretical work on valley splitting in bilayer Si/SiGe structures.  Evidence for exciton condensation has been observed in the $\nu_{\text{T}} = 2$ state for $\frac{\Delta \nu}{\nu_{\text{T}}} < 0.5$.  Achieving independent contacts to both layers in order to perform tunneling conductance,\cite{spielman_2000} Coulomb drag\cite{kellogg_2002} and counterflow measurements\cite{tutuc_2004,yoon_2010} would be required to further confirm this conclusion. Reducing the separation between the bottom quantum well and the bottom gate as well as utilizing larger magnetic fields would enable studies of these effects over a wider density range.  These results highlight the prospects of Si/SiGe bilayers as a platform for valley splitting tuning and exciton condensation, which may have potential applications towards realizing SiGe-based\cite{xu_2020} or Bose-Einstein condensate based \cite{byrnes_2012} qubits.\\

See the supplementary material for details on the methods used to achieve matched density between the layers, how to extract the excitation gaps as well as results from the four cases used in the SP simulations.\\

This work was performed, in part, at the Center for Integrated Nanotechnologies, an Office of Science User Facility operated for the U.S. Department of Energy (DOE) Office of Science. Sandia National Laboratories is a multimission laboratory managed and operated by National Technology \& Engineering Solutions of Sandia, LLC, a wholly owned subsidiary of Honeywell International, Inc., for the U.S. DOE’s National Nuclear Security Administration under contract DE-NA-0003525. The views expressed in the article do not necessarily represent the views of the U.S. DOE or the United States Government. This work was partially supported by the National High Magnetic Field Laboratory. The National High Magnetic Field Laboratory is supported by the National Science Foundation through NSF/DMR-1644779 and the State of Florida. The National Taiwan University (NTU) group is supported by the Ministry of Science and Technology (MOST), Taiwan under contracts MOST-110-2634-F-009-027,  MOST-109-2622-8-002-003, and MOST-110-2218-E-002-030. The authors have no conflicts to disclose. 

\section*{Data Availability Statement}

The data that support the findings of this study are openly available in IR@UF, reference number IR00011549.

\section*{References}
%\nocite{*}
\bibliographystyle{aapmrev4-1}
%\bibliography{APL_refs_v4}% Produces the bibliography via BibTeX.
%merlin.mbs aapmrev4-1.bst 2010-07-25 4.21a (PWD, AO, DPC) hacked

\end{document}